\documentclass[conference]{IEEEtran}
\IEEEoverridecommandlockouts
\usepackage{cite}
\usepackage{amsmath,amssymb,amsfonts}
\usepackage{graphicx}
\usepackage{textcomp}
\usepackage{xcolor}
\usepackage{algpseudocode}
\usepackage{algorithm2e}
\def\BibTeX{{\rm B\kern-.05em{\sc i\kern-.025em b}\kern-.08em
    T\kern-.1667em\lower.7ex\hbox{E}\kern-.125emX}}
\begin{document}

\title{Joint Quantization and Pruning Neural Networks Approach: A Case Study on FSO Receivers}

\author{Mohanad Obeed and Ming Jian\\
Huawei Technologies Canada Co., Ltd., Ottawa, Canada \\
\{mohanad.obeed, ming.jian\}@huawei.com}


%


\maketitle

\begin{abstract}
Towards fast, hardware-efficient, and low-complexity receivers, we propose a compression-aware learning approach and examine it on free-space optical (FSO) receivers for turbulence mitigation. The learning approach jointly quantize, prune, and train a convolutional neural network (CNN). In addition,  we propose to have the CNN weights of power of two values so we replace the multiplication operations bit-shifting operations in every layer that has significant lower computational cost. The compression idea in the proposed approach is that the loss function is updated and both the quantization levels and the pruning limits are optimized in every epoch of training. The compressed CNN is examined for two levels of compression (1-bit and 2-bits) over different FSO systems. The numerical results show that the compression approach provides negligible decrease in performance in case of 1-bit quantization and the same performance in case of 2-bits quantization, compared to the full-precision CNNs. In general,  the proposed IM/DD FSO receivers show better bit-error rate (BER) performance (without the need for channel state information (CSI)) compared to the maximum likelihood (ML) receivers that utilize imperfect CSI when the DL model is compressed whether with 1-bit or 2-bit quantization.


\end{abstract}


\section{Introduction}

Deep Neural Networks (DNNs) have emerged as a powerful tool for tackling a wide range of problems across various domains, including computer vision, natural language processing, wireless communications, and healthcare. These networks are composed of multiple layers of interconnected nodes, each of which processes input data and passes the result to the next layer. DNNs excel at learning complex patterns and relationships within data, making them highly effective for tasks like image recognition, machine translation, and anomaly detection.

However, DNNs often require a substantial amount of computational resources to train and operate. This is due to the large number of parameters involved, which can easily reach tens of millions or even billions. The recent trend towards larger models, such as transformer-based architectures, has further exacerbated this requirement. In the context of wireless communications, where low latency is a critical factor for 6G networks, fast operations are essential \cite{9136588}. This poses a significant challenge for deploying DNNs in such environments, as the computational demands can hinder real-time performance.

To address the computational challenges posed by large DNNs, researchers have explored various techniques to reduce their memory footprint and accelerate their inference process \cite{han2015deep}. Two promising approaches are quantization and pruning \cite{gholami2022survey}. Quantization involves compressing the original network by reducing the number of bits used to represent each weight or activation. By quantizing weights to lower precision (e.g., from 32-bit floating-point to 8-bit integer), the memory footprint and computational cost can be significantly reduced. Pruning, on the other hand, involves identifying and removing parameters that have a negligible impact on the inference output. This can be achieved by analyzing the importance of each parameter and eliminating those that are close to zero. The overarching goal of both quantization and pruning is to strike a balance between model accuracy and efficiency, enabling DNNs to be deployed in resource-constrained environments such as wireless devices.

Several approaches have been proposed to quantize and prune DNNs \cite{jacob2018quantization, aoudia2019towards, gong2014compressing}. Quantization methods are generally classified into two categories: post-training quantization and quantization-aware training. In post-training quantization, weights are quantized after the model has completed training, with no additional retraining required. In contrast, quantization-aware training incorporates quantization into each training iteration, ensuring that the model learns to accommodate the effects of quantized weights throughout the training process. Our focus here is on the quantization-aware training since the proposed algorithm belongs to this category. A common approach of quantization-aware training is to quantize the model parameters  after each gradient update \cite{jacob2018quantization, hubara2016binarized}. The authors of \cite{hubara2016binarized} proposed to binarize the weights so the multiplication operations are mostly replaced by additions and subtractions.  Another common quantization approach is called vector quantization. This approach is based on grouping similar weights and representing each group with a shared quantized value \cite{gong2014compressing}.  The authors of \cite{carreira2017model, aoudia2019towards} utilized the vector quantization along with the concept of quantization-aware training. They proposed to train the model to solve a constrained optimization problem. The approach is to alternate between learning a DNN with continuous
weights and quantizing the weights.

In this paper, we introduce an algorithm for simultaneously pruning and quantizing the weights of neural networks. Our approach involves dynamically adjusting the loss function, the quantization levels, and pruning thresholds at each training epoch. By incorporating quantization constraints into the loss function, we encourage the full-precision weights to converge towards quantized values. Furthermore, we update the quantization levels and pruning limits based on the evolving weight distributions at each epoch, ensuring that the pruning process is adaptive and effective.  This allows for layer-specific quantization and pruning, tailoring the optimization to the unique characteristics of each network layer. Although numerous research papers have reported that 1-bit weight quantization leads to a significant performance degradation \cite{han2015deep, jacob2018quantization, aoudia2019towards}, the proposed approach demonstrates the opposite, showing only a negligible decrease in performance with 1-bit quantization, as shown in numerical results.

The proposed compression algorithm is examined in a free space optical communication (FSO) intensity modulation (IM)/direct detection (DD) (IM/DD)  receivers, where the compressed convolutional neural network (CNN) detects the transmitted symbols for every interval block. FSO communications is a promising technology for 6G backhaul systems, offering higher transmission rates, easier installation, and lower costs compared to fiber optics \cite{obeed2018survey, jeon2023free}. In addition, it can operate with mobile base stations like satellites and drones. However, FSO faces challenges due to atmospheric turbulence, which can cause signal degradation over long distances. This is because the refractive index of the atmosphere fluctuates due to temperature and pressure variations, leading to phase disturbances and signal intensity fluctuations. Deep learning approaches have been proposed in the literature to mitigate the turbulence impact at FSO receivers \cite{10571014, amirabadi2022low, darwesh2020deep}. However, previous DNN models were not optimized for fast execution or reduced memory. Unlike prior work, this paper evaluates a quantized and pruned DNN at the FSO receiver using our proposed compression approach, comparing it with a full-precision DNN. Numerical results show that the compressed model achieves nearly identical performance to full precision: 1-bit quantization yields minimal performance degradation, while 2-bit quantization performs equivalently.
Additionally, the proposed IM/DD FSO receiver shows superior bit-error rate (BER) performance without channel state information (CSI) compared to maximum likelihood (ML) receivers using imperfect CSI, even with 1-bit or 2-bit quantization.




The rest of this paper is organized as follows. The proposed quantization-pruning algorithm is discussed and presented in Section II. Section III shows the considered FSO systems that are used to evaluate the proposed compression approach. In Section IV, we present the simulation results. Finally, our paper is concluded in Section V.









\section{Quantization-Pruning Approach}
In this section, we explain in details the proposed pruning and quantization approach that can be implemented at the training stage for any kind of neural networks. Our approach builds upon existing learning compression techniques (e.g., \cite{carreira2017model, aoudia2019towards}), but incorporates significant modifications to minimize the compression loss and streamline the training process. The main idea is that when we train the model, we alternatively and gradually push the trained weights every epoch to approach the desired quantization levels and modify these levels to match the given weights distribution. The pruning process is considered with assuming that the value zero is always a quantization level. We first train a model to minimize the required loss function $L(\mathbf{w})$ (e.g., mean squared error or binary cross entropy cost function), where $\bf{w}$ are the learnable weights of the neural network (NN). Then we implement Algorithm \ref{Algo1} to quantize and prune the weights at every epoch. Algorithm \ref{Algo1} shows that in epoch $i$, we implement three steps. First, for a given quantized weights $\hat{\mathbf{w}}$ (the initial values of $\hat{\mathbf{w}}$ is assume to be zeros), we train the model to find the full-precision weights $\bf{w}$ that minimizes the loss function given by 
\begin{equation}
    L_i(\mathbf{w}) = L(\mathbf{w}) + \frac{\mu}{2}\lVert \bf{w} - \hat{\mathbf{w}} - \frac{1}{\mu} \bf{\lambda}\rVert^2,
\end{equation}
where $\hat{\bf w}$ are the quantized and pruned weights, $\mu$ is a hyperparameter that increases over iterations according to a predetermined schedule, and $\bf \lambda$ is the Lagrange multiplier estimate that has the same shape of the weights. Second, for a given $\bf{w}$, we use the distribution of the parameters of each layer to find the quantization levels and the pruning limits using Algorithm \ref{Algo2}. Note that we must have $2^b+1$ quantization levels for each layer, where $b$ is the number of quantization bits and we add one, which is the pruning value (i.e., zero).  Third, these quantization levels and the zero value for each layer are used to find the quantized weights that can be obtained by solving the following problem 
\begin{equation}
   \text{arg}\min_{\hat{\bf w}} \lVert \bf w-\hat{\bf w} - \frac{1}{\mu} \bf{\lambda}\rVert^2.
\end{equation}
After we calculate the quntized pruned weights ($\hat{\bf w}$), we update $\bf \lambda$ and $\mu$ as $\bf \lambda = \bf \lambda -\mu(\bf w- \hat{\bf w})$, and $\mu = \mu a^i$. Note that $\mu$ and $a$ are a scalar values.
\SetKwComment{Comment}{/* }{ */}

\RestyleAlgo{ruled}
\begin{algorithm}
\caption{The main quantization and pruning algorithm}\label{Algo1}
$\mathbf{w} \gets \text{ arg } \min_{\mathbf{w}} L(\mathbf{w})$ \; 
$\hat{\bf w} \gets  \bf 0$\;
$\bf \lambda \gets \bf 0$\;
\For{epoch $i = 0,1,..,$} {
  $\bf{w} \gets \text{ arg }\min_{\bf{w}}  L(\bf{w})+ \frac{\mu_i}{2}\lVert \bf{w} - \hat{\mathbf{w}} - \frac{1}{\mu_i} \bf{\lambda_i }\rVert^2$\;
  Obtain the quantization levels and pruning limits using Algorithm \ref{Algo2}\;
  ${\hat{\bf{w}}} \gets \text{arg}\min_{\hat{\bf w}} \lVert \bf w-\hat{\bf w} - \frac{1}{\mu_i} \bf{\lambda_i}\rVert ^2$ \;
  $\bf \lambda_{i+1} \gets \lambda_i -\mu_i(\bf w - \hat{\bf w})$\;
  $\mu_{i+1} = \mu_i a^i$  \Comment*[r]{$a$ is constant}
}
\end{algorithm}

\subsection{Obtaining the Quantization Levels and the Pruning Limits}
In this section, we presents the proposed new algorithm for obtaining the quantization levels and the pruning limits of every layer. This algorithm has to be implemented every epoch or every large number of samples at the training stage. The idea is based on that first dividing the whole weights distribution into areas. Then we adopt the mean of these areas as the initial centers for clustering the weights. The border of these areas are a set of values that must include the zero value (due to pruning condition), the minimum weight value $w_{min}$, the maximum weight value $w_{max}$, and a number of mean values calculated for bigger areas. The total number of areas is $2^b$, while the total number of borders is $2^b+1$, where $b$ is the number of bits used to quantize the weights. For instance, for $b=1$, the borders set is $\{w_{min},w_{mean},w_{max}\}$, where $w_{mean}$ is the mean of the weights ($w_{mean}$ is usually very close to zero since the weights distributions are mainly symmetric around zero). If $w_{min}<0$ and $w_{max}>0$, the initial centers for 1-bit quantization are $\{c_{N},0,c_{P}\}$, where $c_N$ is the mean of the area bordered by $w_{min}$ and $w_{mean}$ and $c_P$ is the means of the of the area that is bordered by $w_{mean}$ and $w_{max}$. For $b=2$ (i.e., 2-bit quantization), we use both the 1-bit borders and the 1-bit initial centers as a border set for the 2-bit quantization. Then the initial centers for the 2-bit quantization are the means of the areas between the borders. In particular, the 2-bits border set is $\{w_{min},c_N,w_{mean}, c_p, w_{max}\}$ and the initial centers are  $\{c_{N_1},c_{N_2},0, c_{P_1}, c_{P_2}\}$, where $c_{N_1}$ is the mean of the weight values that are greater than  $w_{min}$ and less than $c_N$, $c_{N_2}$ is the mean of the weight values that are greater than  $c_{N}$ and less than $w_{mean}$, $c_{P_1}$ is the mean of the weight values that are greater than  $w_{mean}$ and less than $c_P$, and $c_{P_2}$ is the mean of the weight values that are greater than  $c_{P}$ and less than $w_{max}$. In general, for any number of bit quantization $b$, we take the initial centers and areas borders of $b-1$ and adopt them as borders of the areas for $b$-bit quantization.  
We then calculate the means of these areas as the initial centers. 

After finding the mean of the areas, we have now to impose the pruning area, where its mean is zero. The idea is to propose an iterative algorithm where the weights that are closer to zero than the other centers will be associated to the imposed zero area. Imposing the pruning area will push the other areas and their centers apart from the zero value. Hence, we propose to keep associating the weights to their closest center and update the centers based on the associated weights until convergence. This happened while keeping the zero center fixed to guarantee pruned weights.  Algorithm \ref{Algo2} explains the proposed clustering approach.

\RestyleAlgo{ruled}
\begin{algorithm}
\caption{Obtaining the quantization levels and pruning limits for $b$-bit quantization}\label{Algo2}
\For {Layer = $1,...,L$}{
Adopt the initial centers and borders of the ($b-1$)-quantization as the areas' borders of $b$-quantization\;
Find the areas' centers of the given borders\;
\While{(the centers are not converged)}{
  Associate the weights to the given centers\;
  With keeping the zero center fixed, update the centers with the given association\;
    }
Approximate the centers into their closest power of $2$ number\;
}
\end{algorithm}

\subsection{Approximate the Quantization Levels to be with Power of Two}
\label{Approximation}
It's known that addition and subtraction is less computationally complex than multiplication. Multiplying two $K$-bit numbers can involve up to $K$ bit shifts and $K-1$ additions. One way to reduce the computational cost caused by multiplication is to force the model weights to specific values in a well-chosen codebook. For instance, using the codebook $\{-1, 0, 1\}$ simplifies multiplications to merely changing signs or zeroing out.  Another way is to use a codebook consisting of powers of two, which leads to replacing the multiplications by bit shiftings. 

As shown in Alogrithm \ref{Algo2}, we approximate each quantization level in every epoch to be in the form of power of two numbers. To guarantee an accurate approximation, we specifically approximate every quantization level to be in the form of $f2^i + g2^j$, where $f$ and $g$ are either $1$ or $-1$, and $i$ and $j$ are integers. To find the optimal values of $f$, $g$, $i$, and $j$ that minimize the approximation loss, we propose the following: For a given quantization level (float number) $x$, find the value of  $f$ (sign) and $i$ (integer) that minimizes $|f2^i-x|$. Next, for given $f$ and $i$, find $g$ and $j$ that minimizes $(|g2^j-|f2^i-x|)|$. This method ensures the closest approximation to the desired quantization level.

\subsection{Compression and Computation Complexity Analysis}
This paper considers a layer-wise $b$-bit quantization with pruning. In the worst-case scenario, we assume one additional bit is required to represent pruned weights in each layer. To calculate the compression rate for a layer 
$i$, we need 
$b+1$ bits to encode the indices associated with layer 
$i$. Additionally, 17 bits are required to store the quantization levels, represented in the form $f2^i + g2^j$ (where the integer $i$ and $j$ require 16 bits, and the signs $f$ and $g$ require 1 bit each). Assuming a neural network with $P$ parameters distributed across $L$ layers, the compression rate $C_r$ is given by
\begin{equation}
    C_r = \frac{32\times P}{(b+1)P+ 2^b(17L)}.
\end{equation}
For example, in a CNN with 5 layers and 300,000 parameters, the proposed algorithm achieves a compression rate of 16 for 1-bit quantization and 10.6 for 2-bit quantization.

It is also important to evaluate the computation efficiency of the proposed approach compared to the full-precision CNNs. In full-precision CNN, every multiplication requires up to $32$ bit shiftings and $31$ additions \cite{aoudia2019towards}. Assuming a channel dimension of $H\times W \times D$, each value in layer $i$ results from $H\times W \times D$ multiplications and similar number of additions between the channel in the layer $i$ and the activations coming from the previous layer. This means that the full-precision CNN needs  $32 \times H\times W \times D$ bit shiftings and $31 \times H\times W \times D$ to get a one point value in the next layer. 

In contrast, the proposed algorithm reduces each multiplication to 2 bit shifts and 2 additions, resulting in only $2 \times H\times W \times D$ and similar number of additions. This reduces the computational cost by a factor of $16$.

\section{FSO System Model}
We evaluate the proposed joint quantization and pruning approach on a deep learning based FSO receivers that used to detect on-off shift keying (OOK) symbols. We consider two OOK-FSO system models, which are single-input single-output (SISO) and single-input multiple-output (SIMO) models.  The transmitted signal goes through a turbulent channel, with Gaussian noise added to the received signal.  Optical antennas are used to receive the transmitted signal and forward them to one or more photodetectors, that are able to convert the received optical power into an electrical signal. After that, the electrical signal is sampled and sent to a neural network responsible for detecting symbols. We assume the FSO channel remains stable for short periods, known as coherence intervals, each containing $L$ symbols. The correlation from an interval to the next one is assumed to be zero. For the SIMO system, we assume that the optical antennas are sufficiently spaced apart to ensure the channels of the different optical antennas are spatially uncorrelated.

\subsection{Channel Model}
\label{Channel_Model}
 In the literature, a number of statistical models have been used to characterize FSO channels \cite{khalighi2014survey}. However,  the Gamma-Gamma distribution was the most commonly used to model atmospheric turbulent FSO channels.  Gamma-Gamma distribution can be characterized with the following  probability density function (PDF)
\begin{equation}
\label{CH_PDF}
    f(I)=\frac{2(\alpha\beta)^{0.5(\alpha+\beta)}}{\Gamma(\alpha)\Gamma(\beta)}I^{0.5(\alpha+\beta)-1}K_{\alpha-\beta}(2\sqrt{\alpha\beta}); \  I>0
\end{equation}
where $I$ is the atmospheric turbulence intensity, $K(.)$ is a modified Bessel function of the second kind,  $\alpha$ and $\beta$ are the variances of the small and large scale eddies, respectively, and $\Gamma(.)$ is the well-known gamma function. $\alpha$ and $\beta$ can be expressed as follows
\begin{equation}
    \alpha = \bigg[ \exp\bigg(0.49 \sigma^2/(1+1.11\sigma^{12/5})^{7/6}\bigg)-1\bigg]^{-1}
\end{equation}
\begin{equation}
    \beta = \bigg[ \exp\bigg(0.51 \sigma^2/(1+0.69\sigma^{12/5})^{5/6}\bigg)-1\bigg]^{-1}
\end{equation}
where $\sigma^2$ is the Rytov variance, which is given by $\sigma^2 = 1.23C_n^2k^{7/6}z^{11/6}$, where $k=\frac{2\pi}{\lambda}$ is the wave number, and $z$ is the distance between the transmitter and the receiver.

\subsection{Signal Model}
The received signal at the receiver is given by 
\begin{equation}
y = \eta hs+n,
\end{equation}
where $\eta$ is the photodetector responsibility, $h$ is the FSO channel that follows Gamma-Gamma PDF distribution given in (\ref{CH_PDF}), and $n$ is the noise signal that is assumed to be Gaussian with zero mean and $\sigma^2$ variance. 

For SIMO system, there are several versions of the received signals and a combination method is required to minimize the BER. Denote $M$ as the number of optical antennas at the receiver, the received vector can be expressed as 
\begin{equation}
    \mathbf{y} = \eta \mathbf{h}s+\mathbf{n},
\end{equation}
where $\mathbf{h} \in R^{M}$ is the channel vector that consists of independent channels but experience the same turbulence level. In the following, we assume that $\eta = 1$.

\subsection{Deep Learning Based FSO Receiver}
Here, we introduce the considered CNN structures and discuss how we generate and shape the datasets as input and output samples. The CNNs are assumed to be trained, compressed using the proposed quantization-pruning approach, and deployed to the real systems. When we generate the training dataset, we make sure that the dataset spans over a wide range of SNRs and the FSO channel are characterized with different levels of turbulence.


\subsubsection{Generating and Organizing Datasets}
Her, we discuss how the datasets are generated and formed as inputs and outputs for the CNNs in both SISO and SIMO system models. First, since we consider only OOK modulation, the zeros and ones are generated uniformly. The channel models follow a Gamma-Gamma distribution, as detailed in Section \ref{Channel_Model}. Different variances of noise are considered to cover a spectrum of  SNRs.  The channel value is assumed constant for every $L$ symbols, where $L$ is the coherence interval length. The optical antennas in SIMO system are assumed to be sufficiently separated, which means that the channels are spatially uncorrelated. 
 \begin{figure}[t!]
\centering
\includegraphics[scale=0.6]{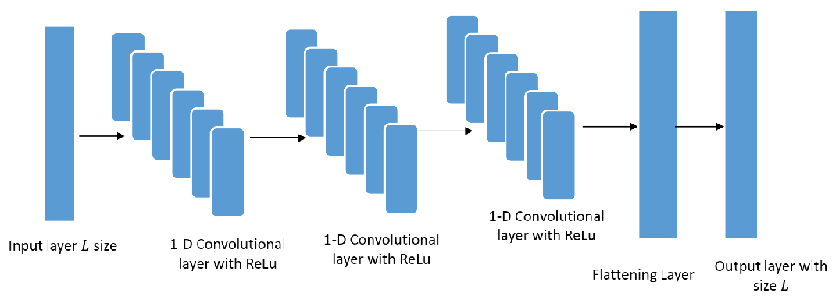}
\caption{CNN structure of SISO-OOK}
\label{fig:CNN}
\end{figure}

\begin{figure}[t!]
\centering
\includegraphics[scale=0.6]{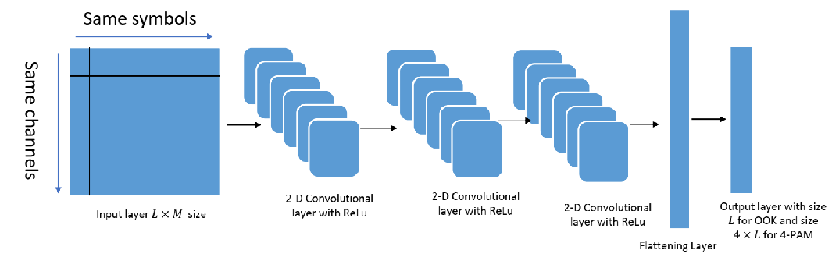}
\caption{CNN structure of SIMO-OOK}
\label{fig:CNN2}
\end{figure}

For SISO system, we organize the input samples as a vector with length $L$. In particular, the input vector $\ell$ with length $L$ to the CNN can be expressed as $\mathbf{x}_{\ell} = h_{\ell} \mathbf{s}  + \mathbf{n}$,  where $h_{\ell}$ is the channel value and it is a scalar, $\mathbf{s} = [s_1, s_2, ... , s_{L}]$ is the symbols vector carried in the coherence interval $\ell$ that are required to be detected, and $\mathbf{n}=[n_1, n_2, ..., n_L]$ is an iid noise vector, Gaussian distributed with zero mean and $\sigma^2$ variance. In a SIMO system with $M$ receiving antennas, different versions of the transmitted symbols are received, each with distinct channels and noise. Since each antenna’s channel remains constant over $L$ symbols, we structure input samples as 
 matrices with size $M \times L$, where the signals in each row characterized with the same channel over time and the signals in each column hold the symbol due to the $M$  antennas. 

\subsubsection{CNN Structure}
This paper considers two CNN designs for SISO and SIMO FSO systems.

For the SISO system (illustrated in Fig. \ref{fig:CNN}), the CNN processes an input of size 
$L$ with three one-dimensional convolutional layers containing 32, 64, and 128 filters, each followed by a ReLU activation. The output of the third layer is flattened and passed to a dense layer of size 
$L$, with a final Sigmoid activation to produce the probability of symbol one.

In the SIMO system (shown in Fig. \ref{fig:CNN2}), input samples are structured as matrices of size $L\times M$. This CNN uses three two-dimensional convolutional layers with the same filter sizes as in the SISO model. The output of the third layer is flattened and connected to a dense layer with a Sigmoid activation for symbol probability estimation.

We use binary cross-entropy as the loss function and the ADAM optimizer to update the trainable variables.



\section{Simulation Results}

\begin{figure}[t!]
\centering
\includegraphics[scale=0.53]{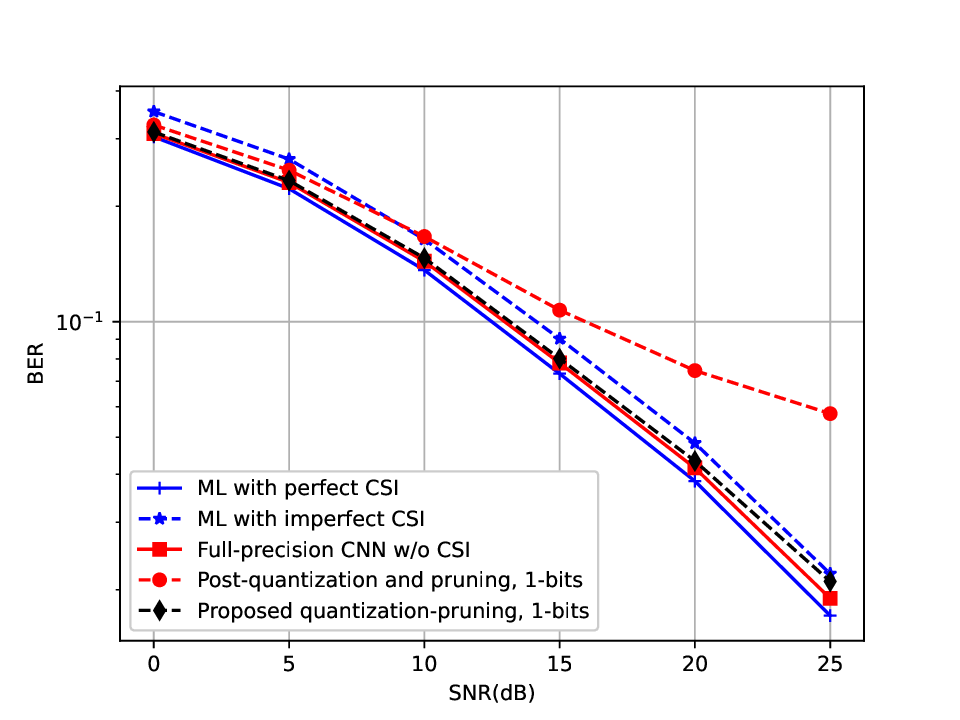}
\caption{BER versus SNR(dB) in SISO-OOK system with 1-bit quantization}
\label{fig:SISO-1bt}
\end{figure}

This section evaluates the proposed quantization-pruning approach over the deep learning based FSO receivers. We evaluate the proposed quantization-pruning approach and DL-based FSO receiver against two sets of baselines. For quantization and pruning, the baselines include a full-precision CNN model and a model that undergoes quantization and pruning post-training. For the FSO receiver, the baselines are ML with perfect CSI and ML with imperfect CSI. The ideal yet impractical baselines (ML with perfect CSI and full-precision CNN) indicate how close the proposed methods are to optimal performance, while the practical baselines (ML with imperfect CSI and post-quantization-pruning) demonstrate the performance gains achieved by our proposed approaches. Note that the full-precision and the compressed CNNs do not use CSI information for symbols detection. The parameters in Algorithm \ref{Algo1}  are initially set as: $a=1.008$, $\mu_0 = 10^{-3}$, $\bf \lambda_0 = \bf 0$, epoch size is $30000$, and the number of epochs is $30$. In channel modeling, we consider moderate turbulence level (i.e., $\alpha = 4$, $\beta = 1.9$). The CNNs are assessed with a coherence interval $L=10$. The number of optical antennas in SIMO systems is assumed to be $M=8$. In all figures, we examine the methods performance in terms of BER with different values of SNRs. 
 \begin{figure}[t!]
\centering
\includegraphics[scale=0.53]{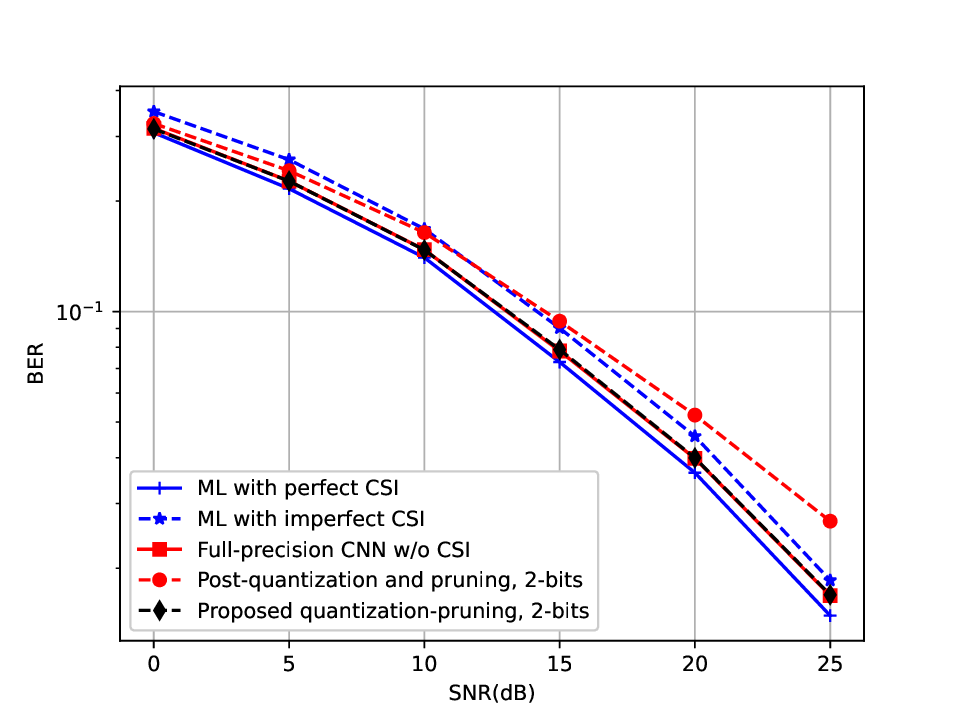}
\caption{BER versus SNR(dB) in SISO-OOK system with 2-bit quantization}
\label{fig:SISO-2bt}
\end{figure}

Fig. \ref{fig:SISO-1bt} shows the performance of the 1-bit quantization and pruning approach of the full precision CNN (32-bits) in SISO-FSO receiver. Note that $b$-bit means $b$ bits used for quantization, while 1-bit is used for pruning. The quantization and pruning is implemented for every layer in the CNN. Hence, in 1-bit quantization, we have only three possible values for every layer, and these values must be 0 and the others with the form $f2^i + g2^j$ as shown in Section \ref{Approximation}. Fig. \ref{fig:SISO-1bt} shows that the 1-bit quantization provides a very close performance to the full-precision CNN. The figure also shows that the quantized and pruned CNN with 1-bit quantization and without utilizing CSI information provides lower BER compared to the ML with imperfect CSI method over all possible values of SNR.

Fig. \ref{fig:SISO-2bt} shows that 2-bit quantization is sufficient to provide the same performance as the full-precision CNN provides. In 2-bit quantization, the possible parameters values in each layer is $2^2+1$. In Figures \ref{fig:SISO-1bt} and \ref{fig:SISO-2bt} we can see the significant gap between the post-quantization and pruning approach and the proposed quantization and pruning algorithm (Algorithm \ref{Algo1}). This implies that selecting the best quantization levels and adapting the loss function every epoch provide significant improvement in reducing the compression loss.

\begin{figure}[t!]
\centering
\includegraphics[scale=0.53]{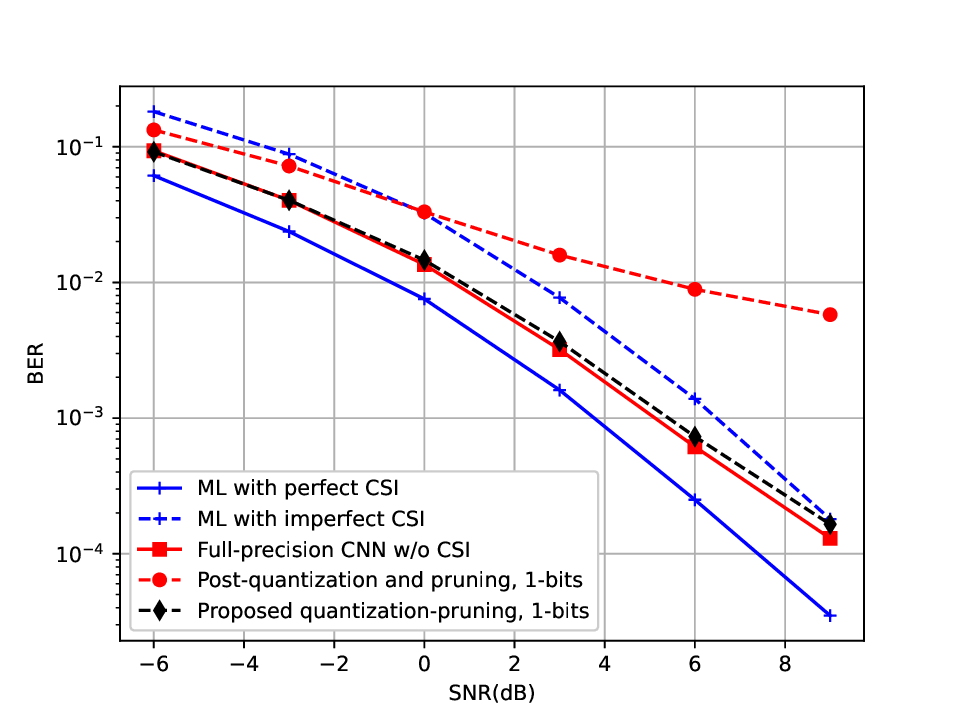}
\caption{BER versus SNR(dB) in SIMO-OOK system with 1-bit quantization}
\label{fig:SIMO-1bt}
\end{figure}

\begin{figure}[t!]
\centering
\includegraphics[scale=0.53]{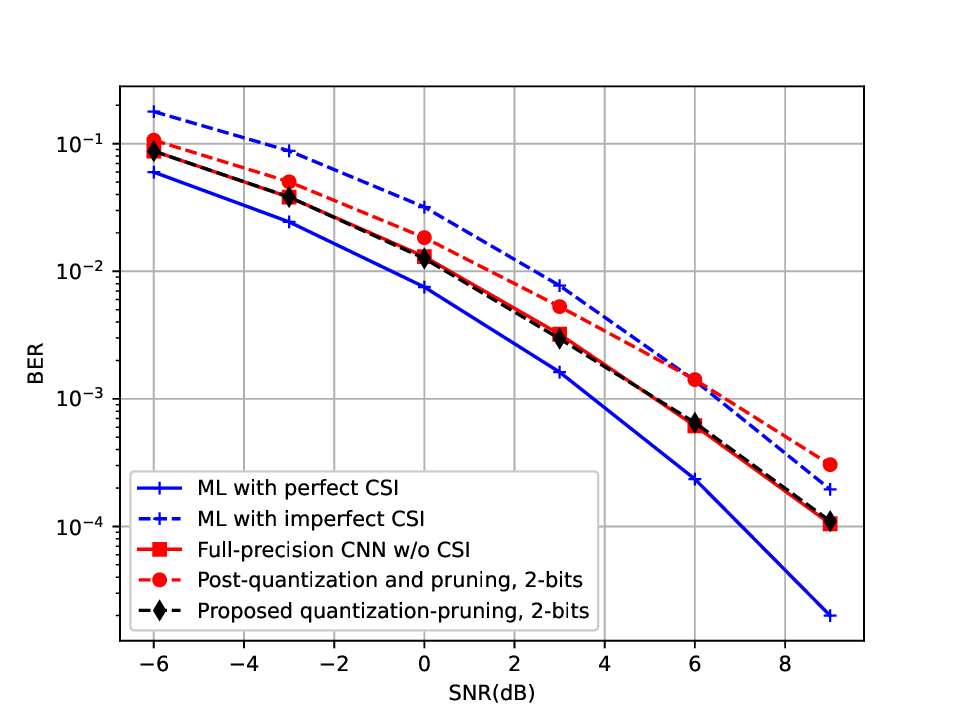}
\caption{BER versus SNR(dB) in SIMO-OOK system with 2-bit quantization}
\label{fig:SIMO-2bt}
\end{figure}

In SIMO-FSO system, we also show the impact of quantizing and pruning CNNs on the DL-based FSO receivers in Figs. \ref{fig:SIMO-1bt} and \ref{fig:SIMO-2bt}. Similar to SISO-FSO system, the proposed 1-bit quantization and pruning algorithm provides a negligible loss compared to the 32-bit CNN as shown in Fig. \ref{fig:SIMO-1bt}. Similarly, Fig. \ref{fig:SIMO-2bt} shows that the 2-bit  provides the same performance as the full-precision CNN without any loss. 

In general, all the figures show that the DL based FSO receivers are better than the ML receiver that additionally need CSI to detect symbols, whether these CNNs are quantized with 1-bit, 2-bit, or kept fully precised. 

\section{Conclusion}
In this paper, we proposed a quantization-pruning algorithm for neural networks. We evaluated this compression algorithm on DL-based FSO receivers, comparing it with both full-precision neural networks and traditional ML FSO receivers. The proposed algorithm achieves performance equivalent to full-precision neural networks with 2-bit quantization, and, in the worst case, shows only a negligible performance drop with 1-bit quantization. Compared to traditional post-training quantization and pruning approaches, the proposed algorithm significantly reduces the BER across varying SNR levels.

\bibliographystyle{ieeetr}
\bibliography{Ref.bib}

\end{document}